# Recent advances in exciton based quantum information processing in quantum dot nanostructures.

Hubert J. Krenner[1], Stefan Stufler[2], Matthias Sabathil[1], Emily C. Clark[1], Patrick Ester[2], Max Bichler[1], Gerhard Abstreiter[1], Jonathan J. Finley[1] and Artur Zrenner[2]

[1]*Physik Department and Walter Schottky Institut, Technische Universität München, Am Coulombwall 3, 85748 Garching, Germany*
[2]*Universität Paderborn, Warburger Str. 100, D-33098 Paderborn, Germany*

**ABSTRACT**

Recent experimental developments in the field of semiconductor quantum dot spectroscopy will be discussed. First we report about single quantum dot exciton two-level systems and their coherent properties in terms of single qubit manipulations. In the second part we report on coherent quantum coupling in a prototype "two-qubit" system consisting of a vertically stacked pair of quantum dots. The interaction can be tuned in such quantum dot molecule devices using an applied voltage as external parameter.

## Introduction

The use of coherent phenomena for the implementation of quantum information technology is expected to provide significant scope for advanced developments in the future [1]. Semiconductor quantum dot (QDs) nanostructures are artificial atoms and hence suitable entities to implement arrays of qubits for solid state based quantum information processing. One possible approach is the use of excitonic excitations in the ground state of a QD as basis for a two-level system. Recently coherent population oscillations, so called Rabi oscillations [2], have been demonstrated in the exciton population of single QDs [3, 4, 5, 6, 7, 8]. Low temperature dephasing times for excitons in self-assembled QDs have been shown to exceed several hundred ps, allowing for large numbers of coherent manipulations with ps pulses before decoherence occurs.[9, 10]

In the present paper, we summarise recent experimental and theoretical developments in the field of semiconductor based quantum information research. The paper is organized as follows: In the first section we report about single QD exciton two-level systems, their coherent properties in terms of a single qubit manipulations, and quantum interference. In the second section, we demonstrate coherent quantum coupling in a prototype "two-qubit" system consisting of a vertically stacked pair of QDs. Furthermore, we show that the qubit-qubit interaction can be tuned using an external parameter. Such controlled coupling is required to perform conditional quantum operations using an intrinsically scalable system and, as such, may be a vital resource for the future implementation of quantum algorithms on the basis of semiconductor nanostructures. Finally, we present a brief outlook in section 3.0.

## Section 1. Single quantum dot photo diodes

The experimental results presented here have been obtained from the ground states of single self-assembled $In_{0.5}Ga_{0.5}As$ QDs. Within this section we further concentrate on n-i-Schottky diodes grown by molecular beam epitaxy on a (100)-oriented $n^+$-GaAs substrate. While based on a conventional diode structure, here a GaAs n-i-Schottky structure, the only optically active part is a single self-assembled $In_{0.5}Ga_{0.5}As$ QD contained in the intrinsic layer of the diode (see Figure 1.1a). The QDs are embedded in a 360 nm thick intrinsic GaAs-layer, 40 nm above the n-doped GaAs back contact. A semitransparent Schottky contact is provided by a 5-nm-thick titanium layer. The optical selection of a single QD is done by shadow masks with apertures from 100 to 500 nm, which are prepared by electron beam lithography from a 80-nm-thick aluminium layer (see reference [11]). For resonant excitation we use a tunable Ti:Sapphire laser, which is focused on the sample by a NA = 0.75 microscope objective. All experiments were carried out at 4.2 K.

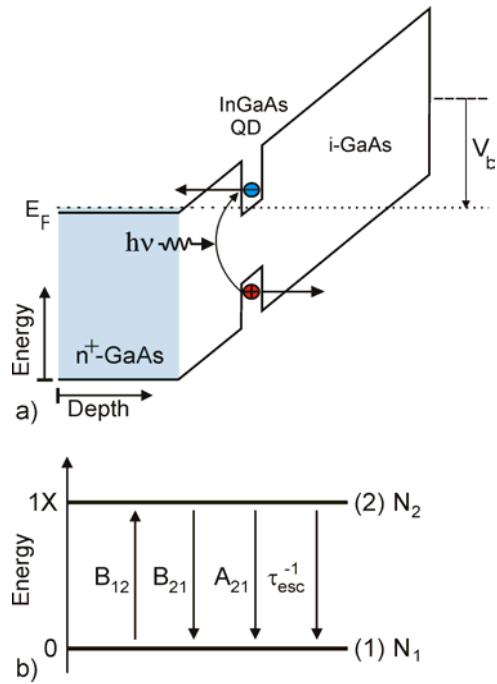

Figure 1.1. (a) Schematic band diagram of a single QD Schottky photodiode for photo current experiments. (b) Fundamental processes in an excitonic 2-level system in the presence of electric field: Transitions are controlled by absorption $B_{12}$, stimulated emission $B_{21}$, spontaneous emission $A_{21}$, and tunneling $\tau_{esc}^{-1}$.

A QD is an artificial atom in a semiconductor, which acts as a protective container for quantized electrons and holes. A single QD photodiode as used here essentially is an exciton two-level systems with electric contacts. In addition to the fundamental optical processes ($B_{12}$, $B_{21}$, $A_{21}$ indicated in Figure 1.1b), substantial tunneling escape $\tau_{esc}^{-1}$ appears at electric fields beyond about 35 kV/cm. The diode arrangement allows for photocurrent (PC) detection, a very sensitive and, as a matter of fact, quantitative way to determine the excitonic occupancy of the two-level system.

## 1.1. Ultra narrow linewidth

Recently PC experiments on QD ensembles have given insight into the mechanisms and time scales of carrier capture, redistribution and escape processes [12, 13, 14, 15]. Furthermore PC experiments on single self assembled QDs have been performed, showing the discrete absorption characteristics of single QDs resulting in sharp spectral features [11]. By use of the quantum confined Stark effect the quantum dot ground state and hence the Eigenenergy of the two-level system can be nicely tuned in energy (see figure 1.2).

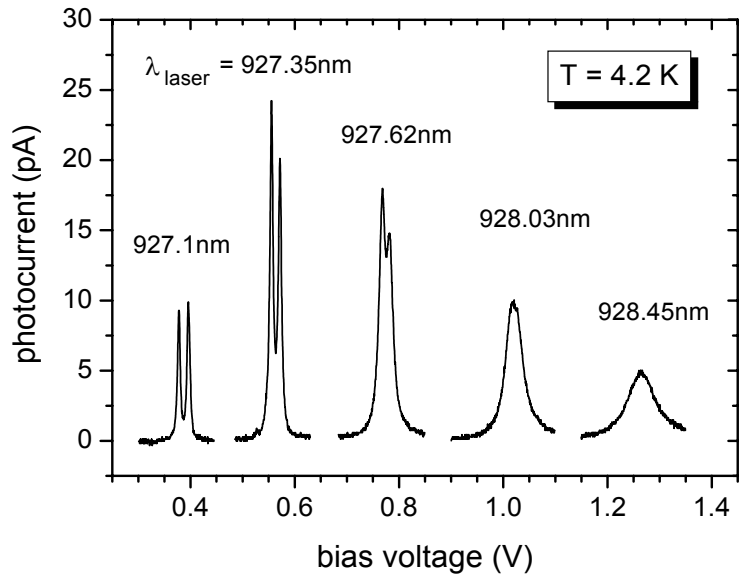

Figure 1.2. PC resonance for various excitation wavelengths versus bias voltage. At low bias the fine structure splitting is fully resolved, at higher bias the linewidth is increased due to fast tunneling.

If the electric field is increased the transition energies of the QD shows a red shift. Within a limited range a PC spectrum can be obtained by a sweep of the bias voltage at fixed laser wavelength. Figure 1.2 shows a number of spectra all representing the same QD state, namely the ground state one exciton (1X) resonance. The excitation wavelength was slightly increased for each spectrum, resulting in a shift of the resonance towards higher bias voltages.

All spectra in figure 1.2 were taken at the same excitation power of approximately 65 nW. One immediately recognizes a notable increase in linewidth at higher bias voltage levels. In the regime of low bias voltages two sharp peaks are clearly distinguishable whereas at high voltages only one broad peak can be observed. A detailed measurement of the Stark effect gives us a conversion of voltage into energy scales with a relative uncertainty of less than 3%. Thus we are able to infer an increase in linewidth from 9 µeV at 0.4 V to about 150 µeV at 1.25 V. This increase corresponds to a higher tunneling probability and therefore shorter lifetime of the investigated 1X state with increasing electric fields. The doublet line structure visible at low voltages can be further investigated by control of the polarization of the excitation beam. On rotating the orientation of linear polarization each peak can be clearly suppressed with respect to the other. This can be explained by a slight shape asymmetry, present in almost any self assembled QDs, resulting in an energy splitting between wave functions oriented along or perpendicular to the elongation axis [16, 17]. We observe here an energy difference between both levels of about 30 μeV. The fact that this splitting is clearly visible underlines the high spectral resolution of our experiment.

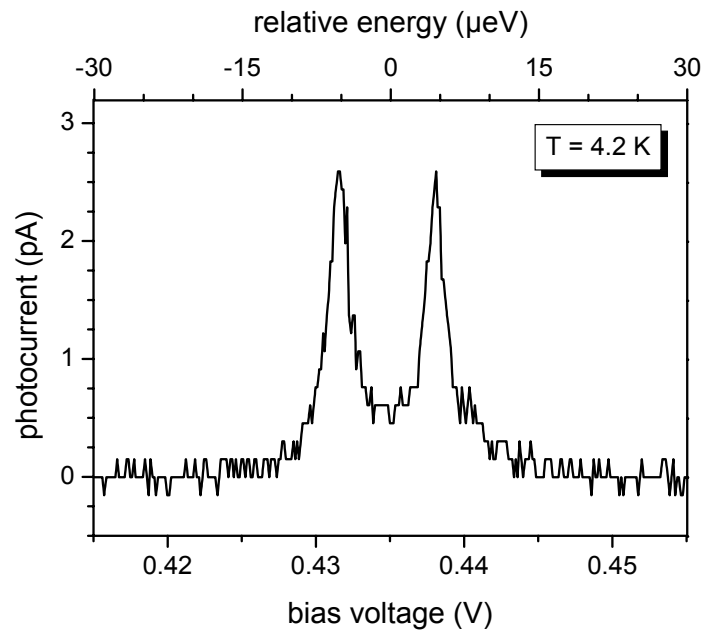

Figure 1.3. PC resonance at fixed wavelength: The bias voltage can be converted to an energy scale via the Stark effect, as shown on the upper axis. A linewidth below 3.5μeV and an asymmetry splitting of 11 μeV are observed for this QD.

We actually expect the resolution to be only limited by the linewidth of the laser used for excitation. Resonant absorption spectroscopy therefore generally is a very capable method for line shape and fine structure analysis of QDs [18]. Due to improvements in sample design we were able to substantially reduce the linewidth of the QD resonance. The spacing between QDs and $n^+$-back contact is 40nm in the present case as compared to 20 nm in the experiments reported in Ref. [19]. As a consequence, we think, Coulomb interactions between QD states and fluctuating background charges have been substantially reduced, which otherwise cause considerable dephasing and therefore an increase in linewidth. In Figure 1.3 we show our so far best experimental result with a directly measured linewidth below 3.5 μeV and a fine structure splitting of 11 μeV.

## 1.2. Non-linear saturation

Another feature of the spectra displayed in Figure 1.2 is the variation of their respective peak height. The decrease at high voltages can be explained simply by the fact that one would expect the integrated signal to be fairly constant rather than its height. The decrease at 0.4 V has to be explained otherwise. Here the tunneling time increases to values similar to the radiative recombination time, resulting in a quenching of PC signal. At even lower bias voltages optical recombination becomes the dominant process and hardly any PC is measurable.

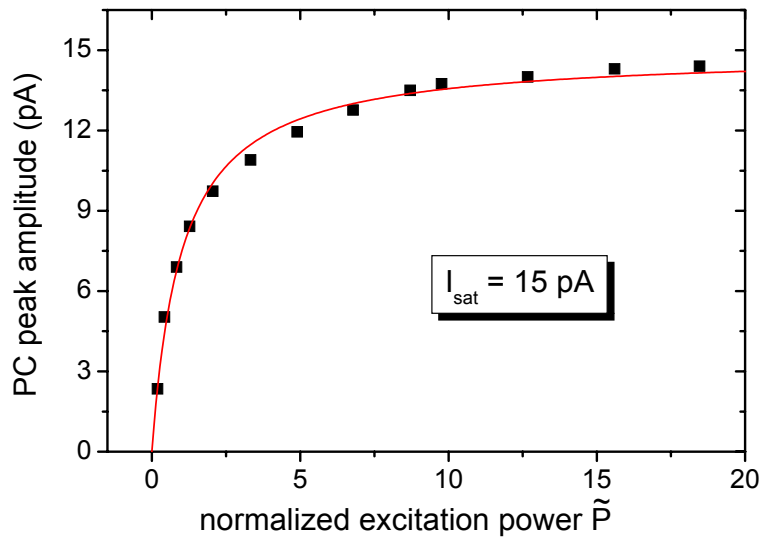

Figure 1.4. Analysis of the PC peak height (bias voltage 0.4 V), showing a saturation at high excitation power. Normalized power levels are displayed, according to the theoretical model described in the text.

In addition a power dependent bleaching of the absorption also contributes to a reduction of PC peak height. This behaviour has been investigated in detail earlier [19] and will be outlined here briefly in the following. For the analysis we chose a bias voltage range where the tunneling time is long enough to allow for narrow linewidths but short enough so that optical recombination only plays a minor role. Within this range a series of PC spectra has been recorded for varying laser intensities. Each spectrum has been fitted by two Lorentzian lines where data points and fit typically show a correlation of 99.95%. For further analysis the arithmetic mean of the two peak heights is used in order to reduce complexity and to minimize random fluctuations. In figure 1.4 we show such an analysis of peak height versus excitation power for a bias voltage of 0.4 V. A clearly nonlinear power dependence is observed, resulting in a saturation of PC peak amplitude at high excitation. The saturation curve can be described by the following equation [19]: $I = I_{sat} \times \widetilde{P}/(\widetilde{P}+1)$ .

Here $I$ denotes the photocurrent peak amplitude, $I_{sat}$ its saturation value and $\widetilde{P}$ corresponds to the normalized excitation power [20]. The physical content of the PC saturation value can be derived fairly easily: If the QD is already occupied by one exciton no further absorption can take place due to a renormalization of energy levels [21, 22] caused by few particle interactions. This also holds true if the QD is occupied only by one carrier. Under the applied conditions the first tunneling process happens fairly fast whereas the tunneling time $\tau_{slow}$ of the slower carrier can even exceed the radiative lifetime. The observed saturation value then is given by $I_{sat} = e/2\tau_{slow}$ [20], where $e$ is the elementary charge. The evaluation of the measured data gives a PC saturation value of $I_{sat} = 15$ pA and a tunneling time of $\tau_{slow} \approx 5.3$ ns, respectively.

### 1.3. Power broadening

The saturation behaviour outlined before also has a direct effect on the linewidth of absorption peaks, independent of any other line broadening mechanisms. At exact resonance, i.e. at the center of a PC peak, the absorption naturally comes nearest to its saturation value. Therefore with increasing excitation the increase in absorption or in our case the increase in PC signal is weakest at the center of a peak and comparatively stronger at its sides. This results in a broadening of the absorption line known in literature as power broadening [23,24]. If one has a homogeneously broadened absorption peak with a Lorentzian line shape of width $\Gamma_0$ the power broadened peak again is Lorentzian but with an increased width $\Gamma$ according to $\Gamma = \Gamma_0 \times \sqrt{1+\widetilde{P}}$ ).

In figure 1.5 we show an analysis of linewidth versus excitation power obtained from the same set of data as used in Figure 1.4. The same conversion of excitation power to $\widetilde{P}$-values is used for both diagrams. Note that therefore at a fixed value of $\Gamma_0$ the slope of the fit curve in Figure 1.5 is no free parameter! From the fact that data points and fit still show very good agreement we can conclude that no other power dependent line broadening mechanisms are of any significant role in our measurement. The extrapolation of the linewidth to zero excitation power gives a value of $\Gamma_0 = 4.3$ µeV. If one assumes the first tunneling process to be the main dephasing mechanism this corresponds to an escape time of $\tau_{fast} = 155$ ps.

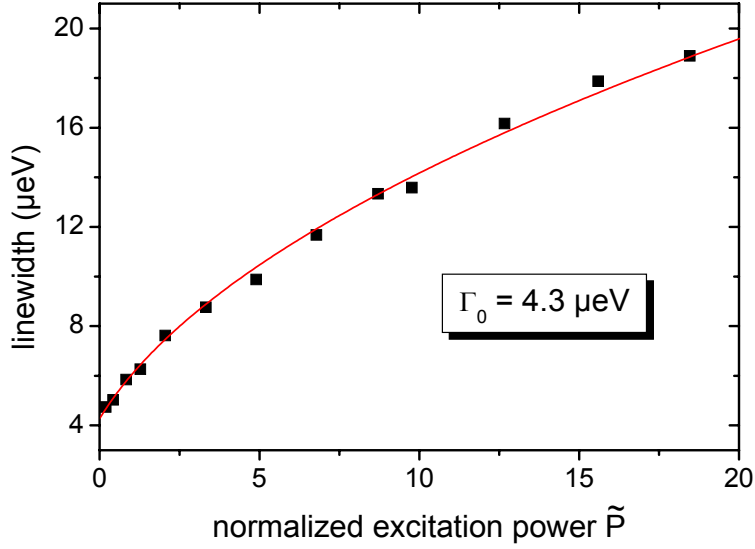

Figure 1.5. Analysis of the power broadening obtained from the same set of spectra as used in figure 1.4. Since the same scaling of the x-axis is applied in both figures, the fit curve contains only one free parameter, which is the linewidth at zero excitation power.

Let us now discuss the results of the analysis shown in Figures 1.4 and 1.5. The saturation value derived in Figure 1.4 gives us a measure for the time it takes the system to go back to its initial state. In genuine two-level systems this would be the lifetime of the excited state, which is usually denoted by $T_1$. In our case this is the escape time of the slower of both photo excited carriers, as discussed above. The linewidth $\Gamma_0$ derived in Figure 1.5 corresponds to the dephasing time of the system, in the context of two-level systems usually denoted as $T_2$. It is important to note that the correct dephasing time can only be derived by an extrapolation to zero power. We performed the according measurements at $V_{bias} = 0.4$ V and infer a linewidth $\Gamma_0$ as low as 4.3 µeV. This reflects a significant increase of the tunneling time towards lower bias voltages. On the other hand a long tunneling time also means that the system needs a long time to come back to its initial state. Saturation and power broadening therefore play an important role even at a comparatively low excitation power. Thus in Figure 1.2 the spectrum at $V_{bias} = 0.4$ V is already notably power broadened (linewidth $\Gamma = 9$ µeV). This shows that particularly in systems with a long lifetime one could easily infer too short dephasing times if power broadening was neglected. The third fit parameter used in the analysis of Figures 1.4 and 1.5 is the scaling of the x-axis. If the intensity of the light field at the spot of the QD was well known one could deduce the oscillator strength of the 1X transition. Due to the near field shadow mask used on our sample this is not possible here. Probably the most important information can be derived from a comparison of the saturation and power broadening analysis. Both sets of data can be fitted well with the same scaling of the x-axis. This means that the observed increase in linewidth is solely a consequence of the saturation behaviour and that dephasing in our system does not increase with increasing excitation power, at least in the regarded range. On the other hand if an analysis of linewidth would yield a different dependence than $\Gamma = \Gamma_0 \times \sqrt{1+\tilde{P}}$ one could get from such a comparison a quantitative measure for power dependent dephasing mechanisms.

## 1.4. Coherent manipulations of a qubit: Rabi oscillations

In the following section we want to focus on the coherent behaviour of the QD. Due to finite dephasing times we use ultrashort laser pulses for excitation. The fundamental experiment in the coherent regime is the observation of Rabi oscillations [2]. The occupancy of the upper level of a two-level system under coherent resonant excitation is given by $\sin^2(\Omega \cdot t/2)$ [23], where the Rabi frequency $\Omega$ is proportional to the square root of the laser intensity and t corresponds to the pulse length. A π-pulse thereby results in a complete inversion of the two-level system. In the context of quantum computing this represents a qubit rotation analogous to the classical NOT operation. We define the pulse area, i.e. the rotation angle $\Theta = \Omega \cdot t$, by adjusting the excitation amplitude rather than the pulse length (see: Figure 1.6b). For the investigated sample a π-pulse typically corresponds to an average laser power on the shadow mask of about 2 μW at a pulse-length of 2.3 ps and a repetition frequency of $f_{Laser}$ = 80 MHz. If the tunnel efficiency of our device was 100 %, any π-pulse would contribute to the PC with one elementary charge, resulting in a maximum value of I = $f_{Laser}$·e = 12.8 pA [5] (see: Figure 1.6a). At low bias voltage however, the tunneling time increases to values similar to the radiative lifetime, which causes a quenching of the PC. At 0.4 V for example, the maximum observed PC is only about 6 pA, as compared to 12 pA at 0.8 V.

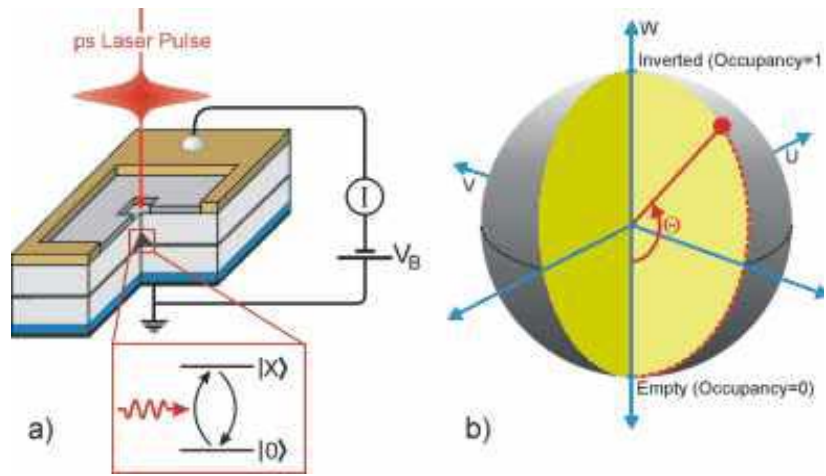

Figure 1.6. a) Coherent excitations of a single QD can be dephased by tunneling and quantitatively measured as PC. b) Bloch sphere representation of a resonant, coherent excitation in a dot. The rotation angle Θ is proportional to the pulse area and to the oscillator strength of the ground state transition.

Figure 1.7 shows the upper level occupancy, reflected in the PC, as a function of the excitation pulse area. At the highest excitation intensities the system undergoes here five full inversions with each laser pulse. The original measurement was corrected for an incoherent background, probably caused by absorption of stray light in wetting layer tail states. As the background is linear in excitation power, it is clearly distinguishable from the oscillatory coherent signal. At π-pulse excitation the incoherent part is about 6 % of the total signal. We use here circular polarized light, in order to avoid biexciton generation not only by spectral separation but also by Pauli blocking [25]. The maximum rotation angle of 6 π as shown here (almost 9π have been demonstrated by us in most recent work) significantly exceeds that of any previously published data on excitonic Rabi oscillations (see for example references [3-8]. Even more important, the observed oscillations only slightly decrease towards high pulse areas, although the excitation power at 6 π is 36 times higher than at π. This is also an experimental proof that the generally observed strong damping at Θ > 1 π is of no principal nature, but usually is caused by sample specifics or by the measurement technique. The data displayed in Figure 1.7 were measured at a bias voltage of 0.7 V, but similar results are obtained in the whole range between 0.4 V and 0.8 V. This is remarkable because the dephasing time of the system varies within this region by about a factor of 6. The damping finally increases at voltage levels where the dephasing time is less than 30 ps, which is about 10 to 15 times the pulse-length.

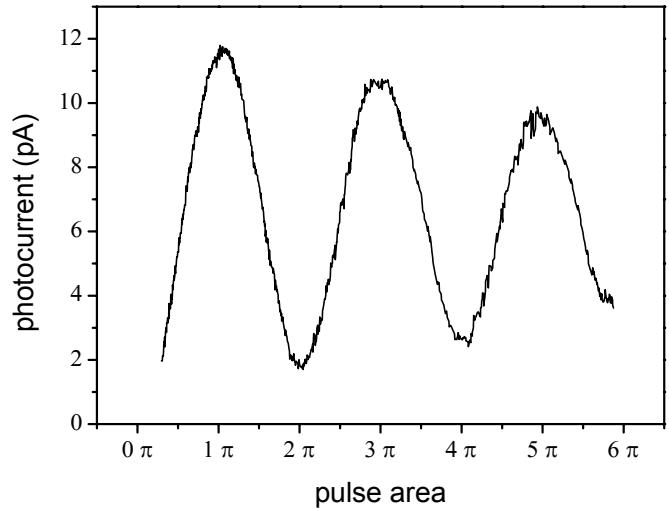

Figure 1.7. Exciton Rabi oscillations for excitation with ps laser pulses (bias voltage 0.6V). The oscillation is only slightly damped towards high pulse areas.

## 1.5. Double pulse experiments: Quantum interference

While the measurement of Rabi oscillations represents the occupancy of a two-level system, we have to perform quantum interference experiments to also gain access to the phase of coherent excitations (see Figure 1.8). First experiments of this kind have been done in the weak excitation regime, i.e. at pulse areas much less than 1 π [9,26]. In order to obtain relevant results on phase coherence with respect to quantum information processing, these experiments have to be extended to the strong excitation regime [4, 6,27].

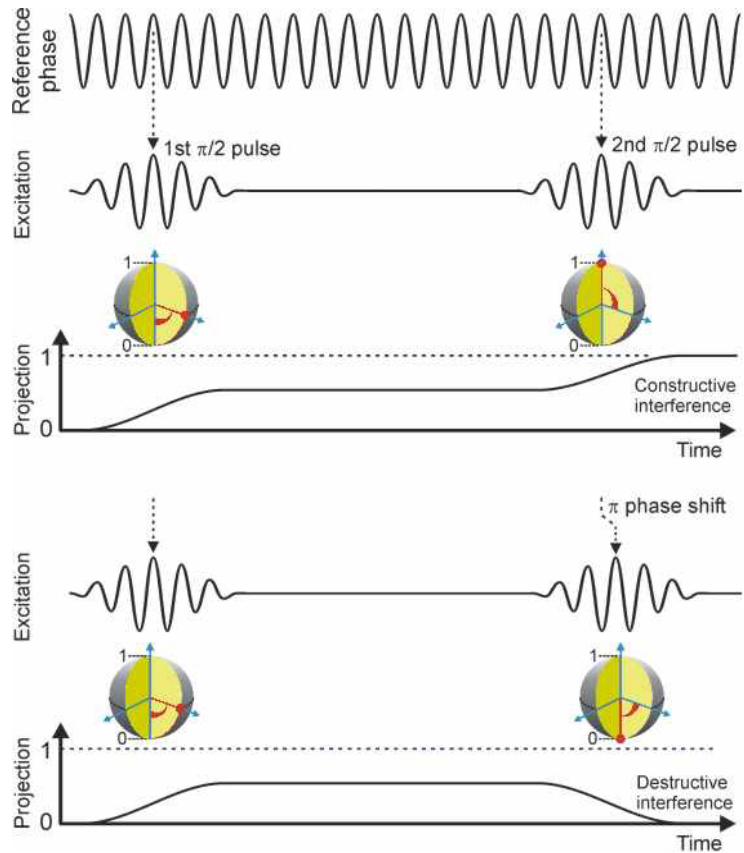

Figure 1.8. Schematic diagrams of a 2-pulse quantum interference experiment on a qubit. The first π/2 laser pulse brings the qubit from |0> into a superposition state and defines thereby also the reference phase in the system. Depending on the phase of the second π/2 pulse, the qubit performs a quantum interference between |0> (phase shift π) and |1> (no phase shift).

We have performed here experiments with π/2 pulses, representing a 1 qubit Hadamard transformation in context of quantum computing [28]. The first pulse thereby creates a coherent superposition of the |0> and |1> state of the QD two-level system. The second pulse then follows with a variable delay in the range of 0 to 1000 ps. The relative phase of the second pulse can be controlled via an additional fine delay with sub-fs resolution. If coherence is maintained, the superposition state is expected to be transferred into the pure |1> or |0> state, depending on whether the two pulses are of the same or opposite phase, respectively. When varying the phase continuously, we observe an oscillation of the PC at the same period as the optical interference at overlapping pulses. The amplitude of these oscillations versus delay time is displayed in Figure 1.9. The data represented has been obtained for linear polarized excitation, for which only one of the asymmetry split levels contributes. A fit to these data points reveals a purely exponential decay at delay times >10ps, corresponding to a dephasing times of $T_2$ = 320ps, 230ps, and 110ps for bias voltages of 0.4 V, 0.48 V, and 0.59V respectively.

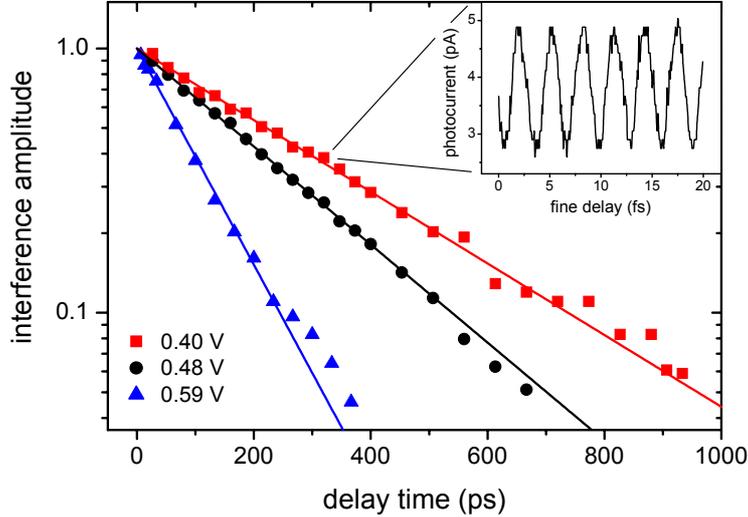

Figure 1.9. Decay of the quantum interference (see inset) versus delay time for different bias voltages. By properly choosing the orientation of the linear polarization, only one resonance from the split ground state was excited (in order to avoid quantum beats).

The analysis of the first few picoseconds is complicated by the fact that an overlap of both pulses to some degree influences the measurement results. We still are able to determine some initial dephasing, though, in the best measurements, this amounts to less than 4 %.

We further are able to compare dephasing times measured by quantum interference with those derived from the linewidth analysis. At low bias and accordingly long tunneling times, however, saturation results in a broadening of the linewidth even at low excitation intensities [20]. We consequently performed a full power broadening analysis with an extrapolation to zero excitation for all measurements up to 0.6 V. At higher bias voltages the PC saturation value is high enough so that the linewidth of single low power spectra will already yield the correct results. The linewidth $\Gamma$ can be converted into a dephasing time $T_2$ via $T_2 = 2\hbar/\Gamma$ [23]. Both sets of data show excellent agreement up to a bias of 0.7 V. At still higher bias we observe quantum beats independent of the choice of polarization and it is therefore difficult to infer a dephasing time.

Resuming the comparison of coherent versus steady state measurements, we get an agreement in several aspects: The ground state linewidth nicely corresponds to the decay time of the quantum interference. The asymmetry induced splitting of energy levels is reflected in the period of quantum beats (not shown here). The polarization at which these effects are suppressed is the same in both measurements. Furthermore the power dependence of different experiments should show some kind of correlation as it is invariably determined by the transition matrix element of the QD. In saturation and power broadening measurements we indeed derive a characteristic dimensionless power level $\widetilde{P} = \Omega^2 T_1 T_2$ [20, 23]. As $T_1$ and $T_2$-times are also obtained in these measurements, we can compare the Rabi frequency $\Omega$ with a direct measurement of Rabi oscillations. In cw-measurements we typically derive a value of $\Omega \approx 0.2$ GHz at a laser power of P = 100 nW. From a comparison of different measurements we get the more general ratio $\Omega/\sqrt{P} = 0.19 \pm 0.3 \, THz/\sqrt{mW}$. In a measurement of Rabi oscillations as shown in Figure 1.7, π-pulse excitation is achieved at an average laser power of 2 μW. A conversion to continuous excitation (assuming sech$^2$ pulses with a FWHM of 2.3 ps at a repetition frequency of 80 MHz) results in a value of $\Omega/\sqrt{P} = 0.25 \pm 0.3 \, THz/\sqrt{mW}$. This is in good agreement with saturation measurements, even though the optical peak power typically differs by five orders of magnitude.

In summary we have performed a whole range of fundamental experiments with respect to two-level systems. All experimental results can be brought down to few basic properties of the investigated single QD, giving evidence for an almost ideal quantum system. We are able here to draw a comparison between complementary experimental methods, so that any indirectly derived parameter can be confirmed by a direct measurement. Furthermore many results mark a major advance in experimentally proven quality, encouraging further work on this kind of quantum system.

In the context of quantum computing the present work demonstrates excellent control over an exciton qubit in a semiconductor QD. In the current experimental setup the ratio of dephasing times versus excitation pulse length would allow for the order of $10^2$ coherent operations. This could be increased on the excitation side by going to shorter laser pulses (see e.g. [29]). In addition any tunnelling related dephasing can be inhibited by applying sufficiently low bias voltage during the coherent manipulation. Electrical readout then would be done by applying a short voltage pulse after the optical qubit rotations have been completed.

**Section 2 Controllable Coupling in Quantum Dot Molecules**

The observation of well resolved Rabi oscillations in the interband optical response of isolated quantum dots (QDs) clearly demonstrates their potential for the realisation of qubits based on electron-hole pair excitations (excitons). As clearly demonstrated in section 1 of the present paper, for the single exciton (=1e+1h) long coherence times approaching the radiative limit are generally observed, much longer than the picosecond duration optical pulses required for quantum state manipulation.[5] The combination of ultrafast optical gating with sensitive schemes for quantitative electrical readout makes such excitonic qubits highly attractive for the implementation of quantum information technologies based on solid state hardware.

A basic requirement for any realistic quantum hardware is the ability to perform *two* qubit operations that, when combined with single qubit rotations, would enable the implementation of arbitrary quantum algorithms.[30] Such conditional quantum operations have already been demonstrated in the restricted basis of one and two exciton states in individual "natural" quantum dots [31] formed from interface fluctuations in quantum wells. However, a major drawback of this approach is that it has little or no prospects for further scalability due to the rapidly diverging complexity of the optical response as the number of carriers in the dot increases.[32,33,34] In the interests of constructing more complex quantum processors using an intrinsically scalable hardware, we have to focus on multiple QD nanostructures. Sophisticated growth techniques such as two-fold cleaved edge overgrowth have already been shown to be suitable for controllably fabricating coupled QDs.[35] A technologically much simpler approach is to vertically stack multiple layers of self-assembled dots separated by thin spacers of the matrix material.[36,37] Over recent years this growth phenomena has been extensively studied for InAs dots embedded within a GaAs matrix.[38,39,40,41] As expected, the structural and electronic properties of the dots are found to depend strongly on the GaAs spacer thickness deposited between each dot layer.[41] For large thicknesses (>55 nm) each layer exhibits properties essentially identical to those of a single layer, with the location of dots in subsequent layers exhibiting no spatial correlation. However, as the GaAs spacer thickness is reduced the positions of dots in different layers become partially correlated. This arises due to the local strain field from one dot layer that gives rise to preferential nucleation sites for dots in subsequently grown layers. For GaAs spacer thicknesses <13 nm the dot positions are fully correlated, with the dots ordered into vertically aligned columns. In addition to this structural ordering there is evidence for electronic coupling between the dots when the GaAs spacer thickness is reduced below ~10 nm.[36] For bi-layer systems, commonly termed QD molecules (QDMs), peak splittings in the photoluminescence spectra have been attributed to the presence of coherent tunnel coupling and the formation of *entangled* e-h pair states.[42,43] Until now the evidence for entanglement is indirect, arising from comparison with simple models and relying on the upper and lower dots being electronically and, therefore, structurally similar. The results of structural microscopy measurements generally reveal that this is not normally the case, with complex variations of dot shape, size and lateral position between the two dot layers.[44] Novel growth techniques such as the In-flush method have been developed to avoid these effects by "shape engineering" the dots in the upper and lower layers.[45,46] Whilst these approaches have been shown to result in structurally similar QDs in each layer of the stack, even if the upper and lower dots could be made *atomistically identical*, their uncoupled electronic structure would differ due to the absence of inversion symmetry along the dot growth axis. In this case, detailed calculations have shown that the degree of entanglement between electron and hole is generally small except for a very small parameter window of dot separation and relative size.[47] The current level of control of the growth of such self-assembled QDMs is insufficient to expect strong intrinsic entanglement between the excitonic states. Additional parameters are required to *tune* the electronic coupling between the upper and lower dots and switch on and off any resulting entanglement. A particularly attractive approach is to tune the electronic coupling by applying static electric fields along the growth direction [14,48,49,50,51] or in the basal plane of the dots.[52] Such approaches have been theoretically demonstrated to facilitate control of the excitonic entanglement [53] and the coherent interaction of the exciton with a light field.[52,53,54]

In this section, we focus on the fabrication and optical properties of such tunable QD-molecules with electric fields applied along the growth direction. For individual molecules we directly observe controlled quantum coupling between different excitonic states. This is manifested by clear anticrossings in the photoluminescence spectrum, from which the strength of the inter-dot coherent coupling is extracted. By comparing our results with realistic calculations of the QD-molecule interband optical spectrum we show that the observed anticrossing occurs between excitons that have predominantly direct and indirect characters with respect to the spatial distribution of the electron and hole wavefunctions. Good quantitative agreement is obtained between experiment and theory and our findings are shown to be very general, similar results having been obtained for more than 25 individual molecules. Statistical analysis of the exciton emission energy, the coupling strength and the electric field at which the excitonic states are tuned into resonance provides very good agreement with our theoretical calculations and expectations regarding the structural properties of our QD-molecules.

The section is organised in the following way: the growth engineering methods employed to realise low density, self-assembled QD molecules with similar electronic structure are introduced in section 2.1. This is followed by a discussion of the excitonic spectrum of QD-molecules and the influence of electric field on the electronic structure (section 2.2). Finally, the devices investigated and results of the single molecule spectroscopy are presented in 2.3.

## 2.1 Fabrication of low density, self-assembled QD-Molecules

The optimisation of self-assembled nanostructure growth techniques in the early to mid 1990 gave significant impetus to the field of single dot spectroscopy. For a recent review of the advances in fabrication and understanding of the physical properties of such self-assembled nanostructures the reader is directed to reference [55]. Experiments such as those discussed in section 1 of the present paper, which are capable of probing the unique quantum mechanical properties of isolated dots, only became possible due to the development of growth techniques capable of engineering the dot transition energy and surface density during growth. For single dot spectroscopy, the QD density has to be sufficiently low in order to use shadow masks with sufficiently high light extraction efficiencies. Furthermore, ground state transition energies higher than ~1240meV are needed to use sensitive silicon-based detectors for emission (photoluminescence - PL) and tuneable Ti-Sapphire lasers for absorption (e.g. photocurrent) spectroscopy. Thus, both the density and emission energy must be optimised to use self-assembled $Ga_{(1-x)}In_xAs$ for experiments in quantum information processing.

Almost all approaches employed until now to realise suitable $Ga_{(1-x)}In_xAs$-GaAs QD material rely on adjusting the $Ga_{(1-x)}In_xAs$ coverage to be just above the critical threshold for self-assembly to obtain a low dot density. Furthermore, control of the In-content enables tuning of the emission energy. We now introduce a widely applied technique to obtain low surface density QDs by interrupting the rotation of the substrate the dot layer growth. This produces a material gradient across the wafer and a position where the coverage is insufficient for QD formation. We will demonstrate how this concept can be readily extended to double layer QD-molecules, which have a high stacking probability, a low surface density and suitable emission energies by carefully adjusting the relative amount of $Ga_{(1-x)}In_xAs$ deposited in the upper and lower dot layers.

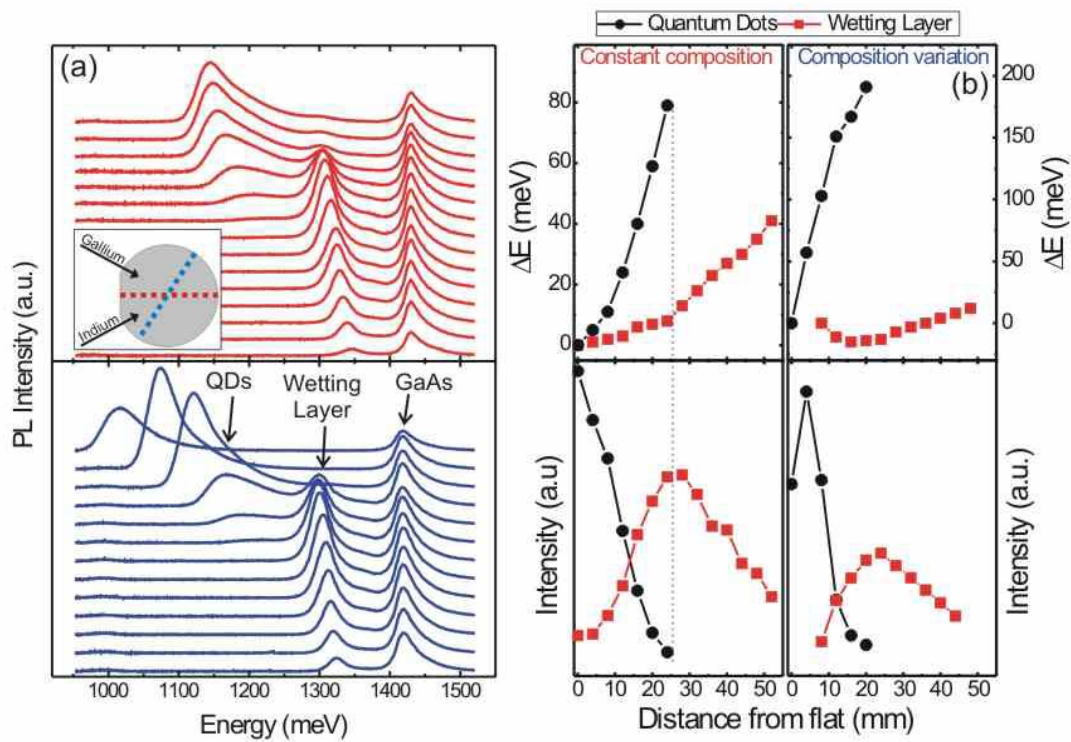

Figure 2. 1 (a) Room temperature PL spectra obtained as a function of position across the wafer for a single dot layer grown by depositing 8ML of $Ga_{0.5}In_{0.5}As$ on GaAs at 530°C without substrate rotation, as discussed in the text. The upper panel (red curves) shows spectra recorded at $\Delta x \sim 4mm$ intervals across the wafer beginning close to the major flat along the direction bisecting the In and Ga cells (red curves) where the In:Ga ratio is kept constant. The lower panel shows similar data recorded along the direction perpendicular to the Ga-cell (blue curves) showing the pronounced influence on the QD emission energy of varying the In : Ga ratio. (b) Peak energy (upper panel) and integrated intensity (lower panel) of the QD (filled circles) and wetting layer (filled squares) along the constant and varying In:Ga ratio directions.

In a first step, single layer samples were fabricated for growth studies by molecular beam epitaxy (MBE) on semi-insulating (001)-GaAs substrates. After growth of a buffer layer the QDs were formed by depositing 8 monolayers (ML) of $Ga_{0.5}In_{0.5}As$ without rotating the substrate at 530°C and a rate of ~0.01nm/s. Dot formation is monitored *in situ* using RHEED and the samples are capped with GaAs for optical experiments. Figure 2.1(a) shows typical room temperature PL recorded from a single layer QD sample at various positions on the wafer under non resonant excitation at λ=632.8nm. The inset shows schematically the Indium (In) and Gallium (Ga) cell geometry in the MBE chamber, the cells being separated by ~33° and the wafer orientated for the dot growth such that the major flat bisects the line between the cells. Along this bisecting line (red dashed line on Figure 2.1) the ratio of the In:Ga flux incident on the wafer is constant whereas perpendicular to the Ga-direction (blue dashed line - Figure 2.1) the stoichiometry varies continuously, becoming increasingly In-rich as one moves further from the Ga cell. Using such approaches it is possible to explore the influence of total material coverage and the In:Ga ratio on the dot density and emission wavelength using only one sample. In both panels of Figure 2.1(a) the distance from the major flat increases from the upper to lower spectrum, but the direction on the wafer corresponds to the constant (varying) In:Ga ratio in the upper (lower) panel. For the PL spectra recorded close to the cells three features can be identified as labelled on the figure: bulk GaAs (~1420meV), the inhomogeneously broadened emission from the QDs (~1000-1250meV) and the two dimensional wetting layer immediately below the QD layer (WL~1320meV). At the position on the wafer closest to the cells near the minor flat, the QD emission dominates the spectra and much weaker emission is observed from the WL. As the wafer is traversed, the QD emission quenches due to reduction of the amount of material deposited and the WL peak gains intensity in an anti-correlated manner. This indicates a shift from a region of the wafer where coverage was sufficient for QD nucleation (3D growth) to a region where only the WL exists (2D growth).

The extracted peak positions and integrated intensities of the QD (circles) and WL (squares) emission are presented in the upper and lower panel of figure 2.1(b) respectively. Along the constant composition direction, both QD and WL peaks initially exhibit a blueshift; their intensities are anti-correlated and vary monotonically. However, ~23mm from the flat the QD signal disappears, the intensity of the WL reaches a maximum and its peak shift rate suddenly increases by a factor of ~3. This position on the wafer marks the point at which a transition of the growth mode from 3D-2D occurs and the dot density is sufficiently low for single dot experiments. After locating this low QD density region on the wafer, the emission energy can be controlled by moving on the wafer in the direction of higher (lower) In:Ga ratio for lower (higher) average emission energy.

The technique introduced above for single QD samples was adapted to realise stacked layers of QD with low surface density and suitable emission energy.[39,56,57,58] For the experiments on isolated pairs of stacked QDs presented below (section 2.3), a high stacking probability has to be guaranteed even for lowest densities. In order to achieve this we started with the growth conditions obtained from the single layer for the lower QD layer. For the upper dot layer, the In-adatom surface diffusion length has to be of the same order as the mean QD separation in the first layer to achieve efficient material diffusion to nucleation sites and the desired high vertical ordering probability. Since the growth temperature required to find the high to low density transition on the wafer is fixed for the lower layer of QDs, we adjusted the material coverage in the second layer to modify the both stacking probability and the size of the upper dot relative to the lower. After growth of the lower dot layer, a *d=7nm* thick GaAs spacer was deposited with the substrate rotation (~6 rotations / s) turned on again. The substrate rotation was then stopped, with the major flat orientated towards the cells as depicted in Figure 2.1, before the upper QD layer was grown. Three different samples were grown for which the material coverage was reduced from 8-6ML in 1ML steps. These samples are denoted 8ML/8ML, 7ML/8ML and 6ML/8ML, respectively.

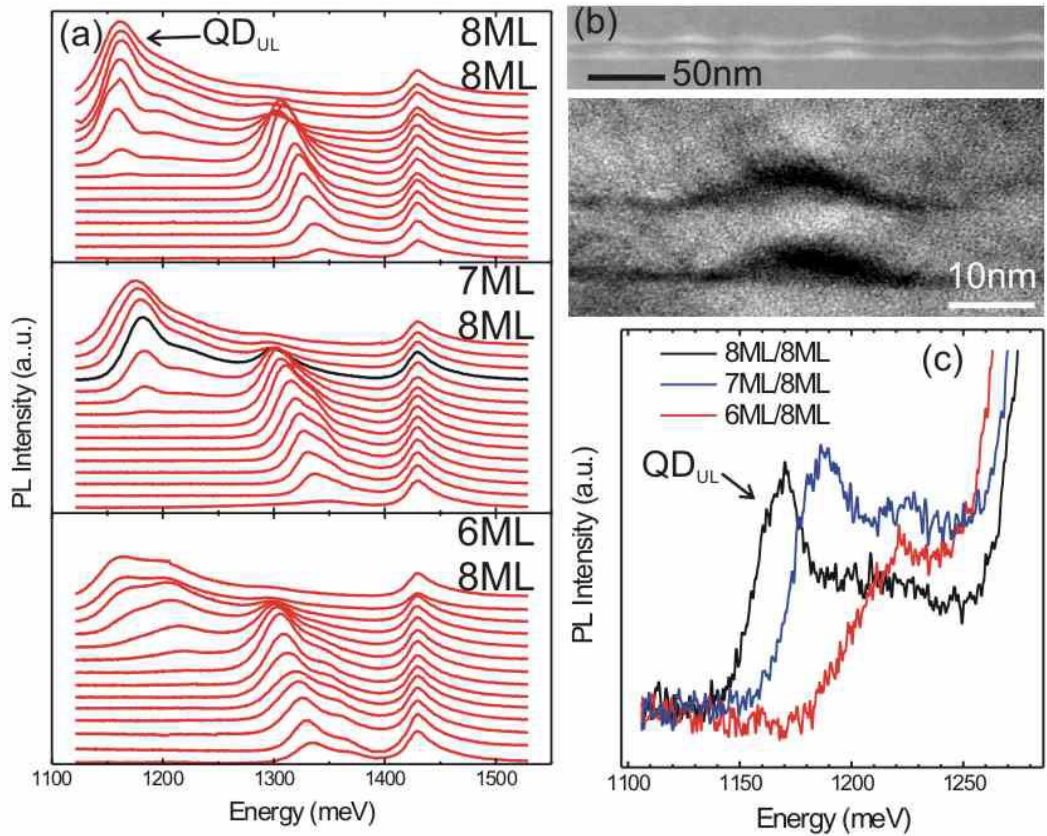

Figure 2. 2 (a) Position dependent PL spectra recorded from the 8ML/8ML (upper panel), 7ML/8ML (middle panel) and 6ML/8ML (lower panel) samples at room temperature. All spectra were recorded along the constant In:Ga ratio direction starting at the major flat (upper spectrum) and moving in $\Delta x=4$mm increments across the wafer. The strongly homogeneous peak arising from the upper QD layer ($QD_{UL}$) can clearly be resolved on a broader background due to the, less homogeneous, lower dot layer. (b) Typical X-TEM microscopy images recorded from the 7ML/8ML sample ~12mm from the major flat (denoted by the black curve on part a). The upper, low resolution image clearly reveals the high stacking probability whilst the lower, high resolution, image shows that the upper and lower dots have a similar size and shape as discussed in the text. (c) Ensemble PL spectra recorded from the low density region of the wafer where single QD-molecule spectroscopy could be performed.

As for the single layer structure, the samples were characterised by room temperature PL recorded at different positions across the wafer along the constant In:Ga ratio line bisecting the cells. The results of these investigations are presented in Figure 2.2(a). Comparing the 8ML(upper)/8ML(lower) sample with the single layer in Figure 2.1(a), an additional narrow feature labelled $QD_{UL}$ on Figure 2.2 is observed on the low energy side of a broader background. The spectral distribution of the background closely resembles the PL spectra obtained from the single layer sample and is attributed to the lower layer of QDs. The energy of $QD_{UL}$ remains almost constant with decreasing coverage and clearly persists over the whole region of the wafer where QD emission is observed. Similarly, for the 7ML(upper) / 8ML(lower) sample

(middle panel – Figure 2.2(a)) this narrower peak is also found to persist over the region of the wafer where QD emission is observed. For the 6ML/8ML sample this feature is only observed in the high coverage region of the sample disappearing ~10mm from the major flat before the point at which a 3D-2D growth transition is reached for the lower layer (~24mm from major flat). Based on these observations, we attribute the narrow peak in the PL data of figure 2.2a to the upper layer of QDs. The prestrained GaAs substrate for the growth of the upper layer seems to improve the dot homogeneity resulting in the observed smaller inhomogeneous linewidth of $QD_{UL}$ when compared with the lower layer. Furthermore, the observation of lower peak energy of $QD_{UL}$ compared to the lower layer is reasonable since the dots formed in the upper layer tend to be larger. This idea is supported by the observed blueshift of the upper layer PL between the 7ML / 8ML and 8ML/8ML samples, arising due to the fact that, after the onset of island formation, more material is available for dot growth producing larger dots in the upper layer.

To confirm that two layers of vertically correlated QDs are formed cross-sectional transmission electron microscopy (X-TEM) was performed on the 8ML/8ML and 7ML/8ML samples at a position on the wafer indicated by the black spectrum in Figure 2.2(a). The upper panel of Figure 2.2 (b) shows an example of low and high resolution images obtained from the on 7ML/8ML sample, in which islands in the two layers are clearly found to be aligned with a stacking probability close to unity. A high resolution bright field image of a single pair of stacked QDs is shown in bottom panel of Figure 2.2(b), showing that the reduced coverage in the upper layer results in the formation of dots with similar size in both layers of the QD-molecule. A statistical evaluation of the high resolution X-TEM data supports this statement, revealing an average base width of ~28nm (~26nm), height of ~5nm (~5nm) and WL thickness of ~1nm (~1nm) for the lower (upper) layer. Our single molecule spectroscopy measurements presented below strongly support this conclusion, since they demonstrate that the exciton transition energies in the upper and lower dots are very similar.

We have presented a readily applicable technique to grow QD-molecules with high stacking probability *and* low surface density by tuning the $Ga_{0.5}In_{0.5}As$ coverage in the upper and lower dot layers with otherwise fixed growth parameters. A reduction of the total $Ga_{0.5}In_{0.5}As$ coverage from 8ML to 7ML for the growth of the upper layer is found to produce structurally similar dots, with comparable transition energies. The growth parameters obtained from this study were used to grow electrically active samples for single QD-molecule spectroscopy presented below in section 2.3.

2.2 Exciton states of QD-Molecules and Quantum Confined Stark Effect

Before presenting the single molecule spectroscopy data we discuss the nature of the single exciton states for our QD-molecules and the influence of static electric field perturbations applied along the QD growth axis. The geometry and composition of the modelled QDMs were based on the X-TEM microscopy investigations discussed above (Figure 2.2(b)) and previous studies of the microscopic In:Ga composition profile of similar, nominally $Ga_{0.5}In_{0.5}As$ dots embedded within GaAs.[59]  A one band effective mass Hamiltonian was used to calculate the single particle states including a realistic treatment of strain and piezoelectric effects. The Coulomb interaction between electron and hole was included self-consistently using the local spin density approximation in order to obtain the single exciton transition energies [60].

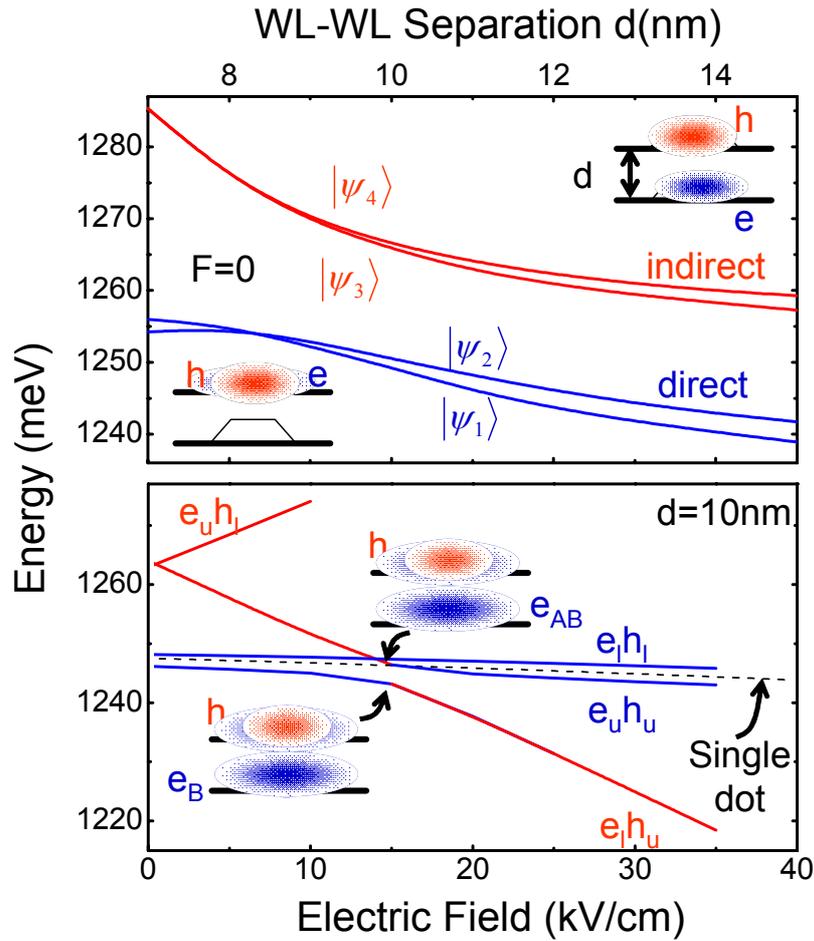

Figure 2. 3 Calculated excitonic states in the model QD molecule consisting of two identical square based truncated pyramids with a height of 5nm, a base length of 20nm situated on top of a 0.5nm thick $Ga_{0.5}In_{0.5}As$ wetting layer. The Indium composition profile of the dots was chosen to have an inverted pyramidal form as deduced from X-STM microscopy performed on GaAs-$Ga_{0.5}In_{0.5}As$ dots grown under similar conditions.[61]  (a) Evolution of excitonic states as a function of the separation $d$ between the wetting layer of the upper and lower QD layers with F=0.  The excitonic states with spatially direct and indirect character discussed in the text can clearly be observed (b) Response of the single excitonic states as a function of static electric field applied parallel to the growth direction.

The calculated energies of the four lowest lying single exciton states of such idealised QD molecules are presented in Figure 2.3(a) as a function of the separation $d$(nm) between the wetting layers of the two dots. Four distinct eigenstates are identified, labelled $|\psi_1\rangle, |\psi_2\rangle, |\psi_3\rangle, |\psi_3\rangle$ in Figure 2.3a.  For d > 9nm, the electron tunnel coupling energy between the two dots is small compared to the *e-h* Coulomb interaction energy and the four exciton states can readily be discussed in terms of a simplified single particle picture.  In this case, $|\psi_1\rangle$ and $|\psi_2\rangle$ are states with predominantly *direct* character for which the electron and hole components of the wavefunction are localised in either the upper ($|\psi_1\rangle = e_U h_U$) or lower ($|\psi_2\rangle = e_L h_L$) dot, respectively. In contrast, $|\psi_3\rangle = e_U h_L$ and $|\psi_3\rangle = e_L h_U$ are states with spatially *indirect* character in which the electron and hole components of the wavefunction are localised in different quantum dots. Obviously, within the framework of this simplified picture $|\psi_3\rangle$ and $|\psi_4\rangle$ are optically dark due to the spatial separation of electron and hole, whilst $|\psi_1\rangle$ and $|\psi_2\rangle$ are optically bright. The eigenenergies of $|\psi_1\rangle$ and $|\psi_2\rangle$ are shifted by ~20meV to lower energy due to the attractive e-h Coulomb interaction for the direct excitons that is absent for the indirect states. In addition, each pair of states in either the direct or indirect branch is generally split by an additional energy of a few *meV*. This effect is not observed using simpler models [51] and arises due to the asymmetry of the confinement potential of the two dots and the long range mutual penetration of the strain field from one dot into the other. For decreasing *d*, the penetration of the strain field becomes much more significant and, additionally, the *electron* component of the exciton wavefunction hybridizes into bonding and antibonding orbitals whilst the hole remains essentially uncoupled due to its much larger effective mass along the growth direction. In this regime of strong tunnel coupling, all the excitonic states exhibit a global blueshift and the separation between the upper and lower pairs of states increases from ~20-35nm as *d* reduces from ~15-7nm. This arises from the combined influence of the reduced electron-hole Coulomb interaction[i] and the widening of the bandgap due to mutual penetration of the strain field from one dot into the other. We note that our calculations presented in Figure 2.3a are in very good quantitative agreement with recent pseudopotential calculations that go beyond our one band model [47], whilst still clearly emphasising the importance of including both strain and Coulomb interactions in the Hamiltonian.

We now discuss the influence of electric field on the states with "indirect" ($e_U h_L$, $e_L h_U$) and "direct" ($e_U h_U$, $e_L h_L$) character for model QDMs similar to those investigated experimentally 7ML / 8ML *d*=10nm, in the weakly coupled regime. The results of our calculations are presented in Figure 2.3b. As discussed above, the two exciton species in weakly coupled QDMs differ in their spatial electron-hole configurations. The energy perturbation due to the quantum confined Stark effect (QCSE) is given by $\Delta E_{QCSE} = -\vec{p}.\vec{F}$, where *p=e.s* is the excitonic dipole due to the

---

[i] The electron component of the wave function becomes delocalised over both dots while hole hybridization is weak

spatial separation (*s*) between the maxima of the electron and hole envelope functions and *F* is the applied electric field. For direct excitons the electron and hole wavefunctions are localized within the same dot and are typically separated by *s<1nm* leading to a small excitonic dipole that is weakly dependent on *F* and is largely insensitive to the dot-dot separation *d*. This gives rise to a *weak*, parabolic Stark shift comparable to the situation observed with single dot layer samples.[14] In contrast, for excitonic states with *indirect* character the electron-hole separation is much larger, of the order of *s=d*, and is largely independent of *F*. This gives rise to a much *stronger* linear quantum confined Stark shift $\Delta E_{QCSE}^{indirect} \approx \pm edF$, the plus / minus sign denoting the opposite orientation of the exciton dipole for $e_U h_L$ and $e_L h_U$, respectively. From these simple considerations, one would expect that application of an electric field along the QD growth axis would result in weak parabolic shifts of both direct exciton branches and much stronger linear shifts of the indirect exciton branches. In particular, $e_L h_U$ is expected to shift to *lower* energy at a rate $d\Delta E_{QCSE}^{indirect}/dF \approx -ed$. Precisely this behaviour can be identified in the calculations of Figure 2.3b where $e_L h_U$ shifts rapidly to lower energy as the electric field increases. As the electric field approaches $F_{crit} \sim 15$kV/cm, $e_L h_U$ comes into resonance with the direct exciton states ($e_U h_U$ and $e_L h_L$) and exhibits a clear anticrossing (crossing) with $e_U h_U$ ($e_L h_L$) – figure 2.3b. This arises since the electron component of the exciton wavefunction hybridizes forming symmetric and anti-symmetric molecular orbitals as shown schematically by the inset on Figure 2.3b.[ii] These two hybridized states are split by a coupling energy of ΔE~5meV for *d=11*nm. Below in section 2.3, we show that precisely this anticrossing between two excitonic states with direct and indirect character is observed by performing spectroscopy on single QDMs. For F>F$_{crit}$ the direct and indirect excitons are tuned out of resonance and the states recover their direct ($e_U h_U$) and indirect ($e_L h_U$) spatial character. We note that, as the electric field passes through F$_{crit}$, the nature of the QDM ground state changes from *direct* to *indirect* as depicted schematically in Figure 2.3b. Based on the preceding discussion, one would expect the indirect exciton to gain oscillator strength as $e_L h_U$ and $e_U h_U$ are tuned into resonance until both branches have similar oscillator strength at F=F$_{crit}$. For F>F$_{crit}$ the ground state loses oscillator strength as it recovers its predominantly indirect

---

[ii] A crossing is observed between $e_L h_U$ and $e_L h_L$ since the hole tunnelling coupling is extremely weak and these two configurations are distinguishable via the location of the hole in the upper or lower dot, respectively.

character and would be expected to vanish in the optical spectrum. We demonstrate below that precisely this behaviour is observed in our single molecule spectroscopy experiments.

2.3 Single QD-Molecule Spectroscopy

In order to test the predictions of the calculations presented in the previous section we performed low temperature PL spectroscopy on single QDMs embedded in n-type Schottky photodiodes using techniques similar to those employed in section 1 of the paper. The epitaxial layer sequence was as follows: firstly, a highly doped $n^+$-back contact was grown followed by 110nm of undoped GaAs. The two layers of nominally $Ga_{0.5}In_{0.5}As$ dots were then deposited separated by a d=10nm thick GaAs spacer and using the 7ML / 8ML growth conditions discussed in section 2.1. After deposition of the QD-molecules the samples were completed with 120 nm of undoped GaAs. As shown schematically in Figure 2.4a, the samples were processed into photodiodes equipped with opaque aluminium shadow masks patterned with ~500nm diameter apertures to isolate single QDMs for optical investigation. A schematic of the device structure is presented in Figure 2.4b. By applying a bias voltage $V_g$ between the n-contact and the Schottky-gate, the axial electric field ($F$) can be tuned from 0 to ~250kV/cm. For low electric fields (F<30kV/cm) the exciton radiative lifetime is shorter than the carrier tunnelling escape time out of the QDMs and photoluminescence experiments could be performed (see section 1).

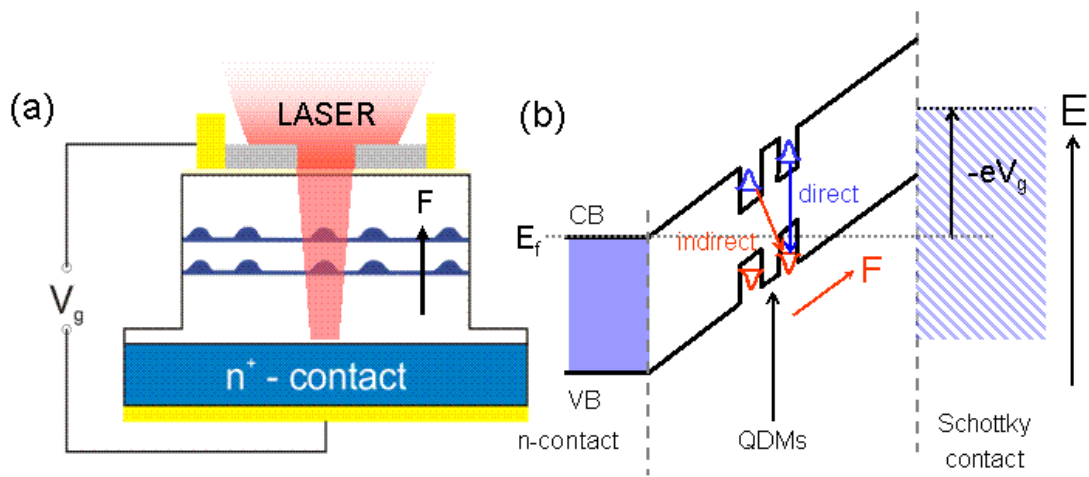

Figure 2. 4 Schematic device structure (a) and band profile (b) of the single molecule Schottky photodiode samples investigated. The band profile shows schematically an applied bias such that F>$F_{crit}$ and X($e_Uh_U$) and X($e_Lh_U$) are out of resonance.

Typical PL-spectra recorded from a single QD-molecule are presented in the main panel of Figure 2.5a as a function of electric field (F) increasing from the upper to lower spectrum. The spectra presented were recorded using a very low excitation power density ($P_{ex}$~2.5Wcm$^{-2}$) to ensure that single exciton species dominate the emission spectra. For the lowest electric fields investigated a number of lines are observed in the emission spectrum distributed over a ~5meV wide window around ~1289meV. Measurements of the intensities of each of these peaks as a function of excitation power density (not shown) demonstrate that the peaks labelled $X(e_L h_L)$ at 1289.2meV and $X(e_U h_U)$ at 1289.8 vary linearly with $P_{ex}$. These features are, therefore, identified as the *single,* optically active exciton transitions in the QDM as discussed in the previous section.[33,34] In contrast, the other peaks labelled 2X and 2X′ are found to exhibit a superlinear power dependence suggesting that they arise from biexciton states in the QDMs. A discussion of these higher order exciton complexes will be presented elsewhere, here we focus only on the single exciton species for comparison with the calculations discussed in section 2.2.

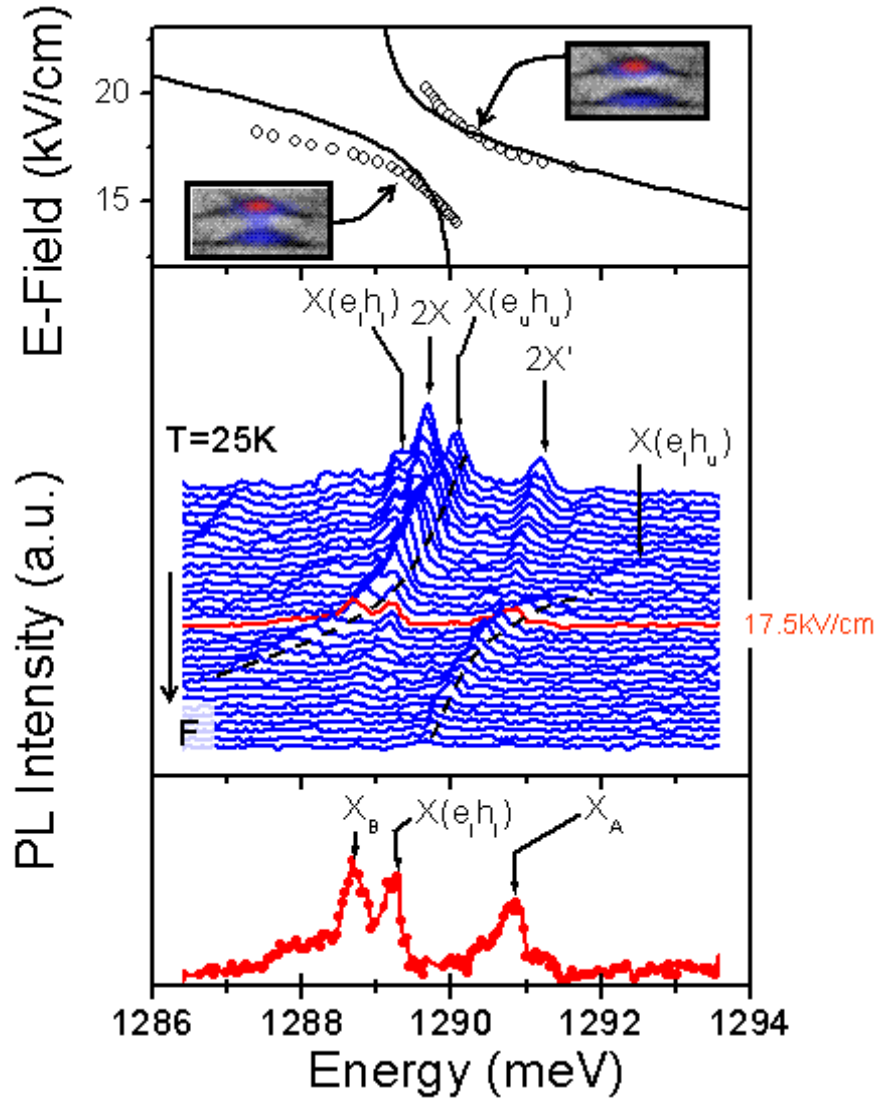

Figure 2. 5 (main panel) Photoluminescence spectra recorded at T=25K from a single QD-molecule under conditions of weak excitation as a function of static electric field. The anticrossing discussed in the text can clearly be observed at an $F_{crit}$~17.5kV/cm. (lower panel) PL spectrum at the anticrossing showing the bonding $X_B=1/\sqrt{2}(X[e_Uh_U]+ X[e_Lh_L])$ and antibonding branches ($X^A=1/\sqrt{2}(X[e_Uh_U]-X[e_Lh_U])$) and the uncoupled exciton $X(e_Lh_L)$. (upper panel) Comparison of peak positions of two excitonic branches and the calculated exciton energies for a model dot with the parameters deduced from the X-TEM measurements of section 2.2 and the calculation method discussed in 2.3.

We begin by discussing the peak $X(e_Uh_U)$. This feature shifts weakly with increasing field until, at a critical field of F~17.5kV/cm denoted by the red spectrum in Figure 2.5, the shift rate suddenly increases and the peak intensity quenches rapidly. Over the same range of electric field, another peak with single exciton power dependence, labelled $X(e_Lh_U)$ is observed at higher energy. This feature shows precisely the opposite behaviour: it is weak and shifts rapidly for F<17.5kV/cm whereafter the shift rate suddenly decreases and the peak gains intensity.[50] These two features clearly anticross at $F_{crit}$=17.5kV/cm as depicted by the dashed lines on Figure 2.5 that serve as a guide to the eye. This behaviour is in good accord with the expectations of section 2.2: $X(e_U h_U)$ exhibits a weak Stark shift corresponding to a direct exciton state, whilst $X(e_L hU)$ shifts more strongly with electric field as expected for an indirect state. At the anticrossing the two observed peaks correspond to the bonding ($X_B = \frac{1}{\sqrt{2}}(X(e_U h_U) + X(e_L h_U))$) and antibonding $X_B = \frac{1}{\sqrt{2}}(X(e_U h_U) - X(e_L h_U))$ molecular like states due to hybridization of the electron component of the exciton wavefunction as discussed in section 2.2. Such anti-crossings are a clear signature of coherent quantum coupling of the excitonic states in the QD-molecule, a prerequisite for the implementation of such systems for quantum information processing. Furthermore, the energy splitting between the states ($2E_{e-e}$) is a direct measure of the quantum coupling strength obtained using a fully optical technique.

The indirect state is expected to have weak oscillator strength away from the anticrossing and, indeed, the amplitude of $X(e_L h_U)$ in the PL spectrum is weak away from the anticrossing as expected. As the anticrossing is approached, the intensity of the two branches becomes similar since both bonding and antibonding states have finite oscillator strength. The overall reduction in the absolute PL intensity of all features arises from field induced suppression of carrier capture and, thus, a reduction of the effective optical pumping density as discussed above.

The peak positions of both the anticrossing branches are plotted in the top panel of Figure 2.5 as a function of $F$. We adapted the structural parameters of the model QDMs discussed in section 2.2 to fit to the experimental data by slightly adjusting the relative size of the dots in the upper and lower layers and the dot-dot separation. The lines plotted on the top panel of Figure 2.5 compare our calculations with the experimentally measured peak positions of $X_A$ and $X_B$ as a function of $F$. For the fitting, the electric field at which the anti-crossing occurs was found to be most sensitive to the relative vertical height of the lower and upper dots ($h_l$ and $h_u$) whilst the anti-crossing energy was mainly determined by the dot-dot separation $d$. The fit shown in Figure 2.5a was obtained using $h_l$=3.5 nm, $h_u$=4 nm and d=12nm to produce fairly good quantitative agreement with the experimental data.

We now turn to the second peak observed in the low field spectra of Figure 2.5 with single exciton character - $X(e_L h_L)$. This feature shifts weakly over the entire range of electric field investigated and clearly does not show an anticrossing with $X(e_L h_U)$. The observation of a weak shift rate identifies this peak as an excitonic state with direct character. Reference to the calculations presented in Figure 2.3b indicate that it arises from the direct exciton state in the *lower* quantum dot $X(e_L h_L)$. Reference to the calculations presented in Figure 2.3 show that $X(e_L h_L)$ and $X(e_U h_U)$ are typically separated by a few meV at F=0, both peaks exhibiting weak Stark shifts characteristic of states with spatially direct character. For the presently investigated molecule $X(e_L h_L)$ and $X(e_U h_U)$ are almost degenerate at zero field with $X(e_L h_L)$ slightly shifted to lower energy ($\Delta E[X(e_U h_U) - X(e_L h_L)] < 0.5$meV) compared to the idealised geometry used for calculations presented in Figure 2.3. This would suggest that the upper and lower dots are similar in size, a statement that is fully consistent with the X-TEM data presented in Figure 2.2 and the model dot geometries required to fit the peak positions of Figure 2.5 (upper panel).

The characteristic form and electric field dependence of the PL spectra discussed above was found to be very reproducible, similar results having been obtained from more than 25 different QD-molecules. We now present the statistical distribution of the measured coupling energy ($2E_{e\text{-}e}$) extracted from the observed anticrossing between $X(e_U h_U)$ and $X(e_L h_U)$, the "centre energy" between these states $E = (X(e_L h_U) + X(e_L h_U))/2$, and the critical electric field $F_{crit}$ at which the anticrossings are observed. These results of these investigations are summarised in Figure 2.6.

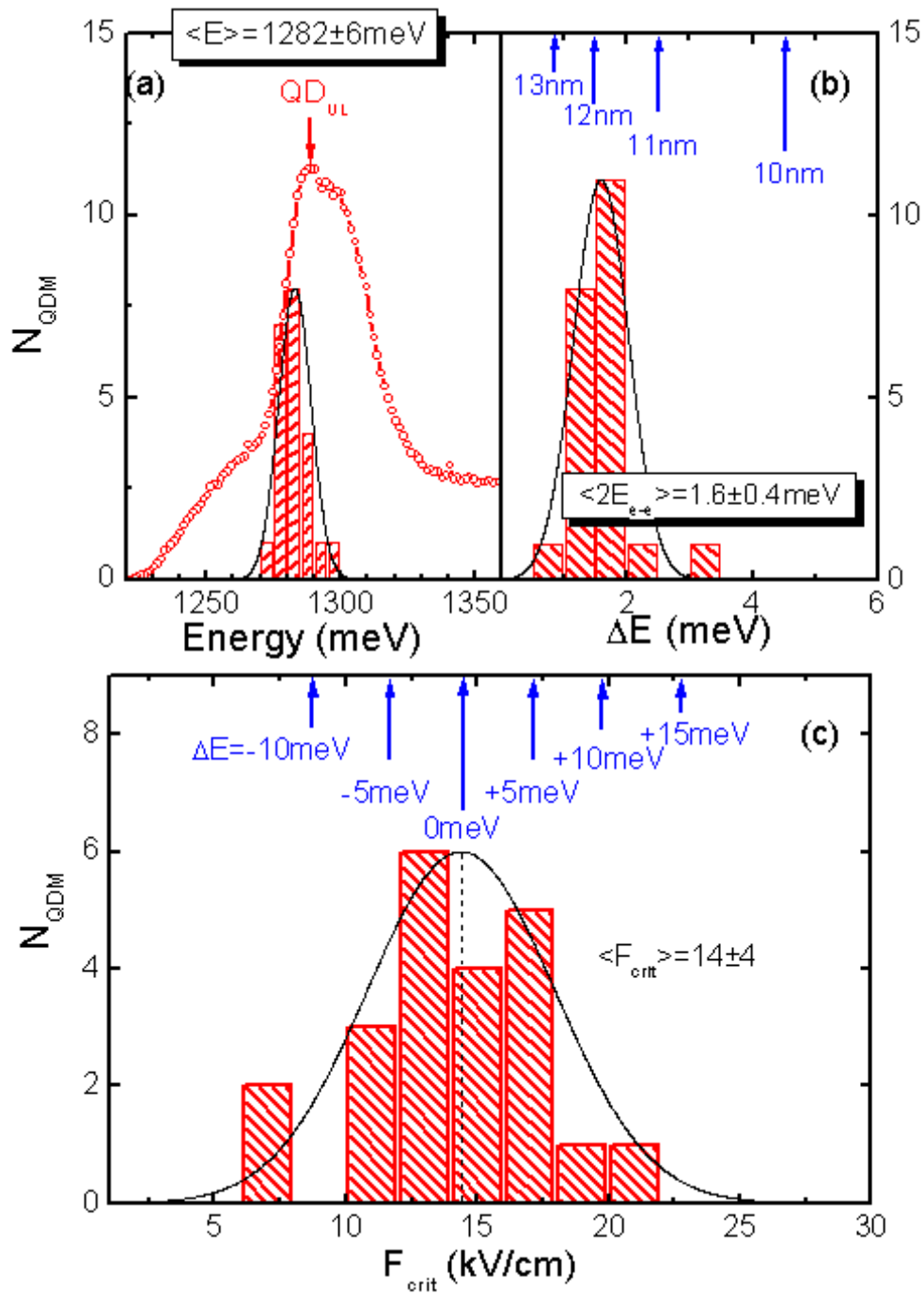

Figure 2. 6 Statistics obtained from single QD-molecule measurements performed on over 20 individual molecules (a) Mean emission energy <E> of observed excitonic anticrossing compared with an ensemble PL spectrum recorded from the same region of the sample. The good agreement between the observed anti-crossings and the PL peak from the upper QD layer confirms that the observed anticrossing occurs between $X(e_U h_U)$ and $X(e_L h_U)$ as discussed in the text and calculated in section 2.3. (b) Splitting energy at the anticrossing <ΔE>, compared with the calculated inter dot separation in the range d=10-13nm. (c) Distribution of the electric field at which the anticrossing occurs ($F_{crit}$) compared to the calculated F=0 energy splitting between the direct exciton states in the lower and upper dots = $E(X(e_L h_L) - X(e_U h_U))$.

Figure 2.6a shows a histogram of the mean energy where anticrossings are observed and the corresponding normal distribution, using a sample bin width of only 1meV.

Most remarkably, we find that anticrossings are only observed in a relatively narrow energy window around 1282±8meV. Comparison of this distribution with the ensemble PL spectrum (open circles – Figure 2.6a) recorded from the position on the wafer used for the single molecule spectroscopy shows that this energy coincides almost exactly with the homogeneous PL peak arising from the upper QD layer ($QD_{UL}$), as discussed in section 2.2. This observation provides additional strong support to the identification of the two states involved in the anticrossing ($X(e_Uh_U)$ and $X(e_Lh_U)$) as being states in which the hole component of the wavefunction is localised in the *upper* dot layer. Furthermore, the ensemble PL spectrum clearly shows that the inhomogeneously broadened emission from the upper and lower QD layers have a strong spectral overlap, lending additional support to the idea that the upper and lower dots have similar electronic structure due to the reduced $Ga_{0.5}In_{0.5}As$ coverage in the upper layer (7ML / 8ML – sample).

The statistical distribution of the quantum coupling strength extracted from the splitting between $X_B$ and $X_A$ is plotted in Figure 2.6b together with the normal distribution, from which the mean coupling strength is measured to be $<2E_{e-e}>$ =1.6±0.4meV. Also shown on the figure by the blue arrows is the calculated quantum coupling strength for a number of dot-dot separations ranging from d=13-10nm, using the model QD-molecule parameters employed to fit the data of figure 2.5. The best agreement with calculation is obtained for d=12±0.5nm, very close to the nominal dot-dot separation of d=10nm for this sample. Experimentally, it appears that the two QD-molecules are more weakly coupled than would be expected for a nominal 10nm separation, for which we calculate $2E_{e-e}$=4.6meV. This discrepancy may reflect, for example, the microscopic In-distribution throughout the QD-molecules which is not expected to be homogeneous.[61] However, it is likely that this minor discrepancy is due to the one-band model used for our electronic structure calculations. Further investigations are currently in progress to study the dependence of $<2E_{e-e}>$ on the nominal dot-dot separation and provide definitive experimental data for more detailed comparison with theory.

The statistical distribution of the field at which anticrossings are observed ($F_{crit}$) is presented in Figure 2.6c. The mean value of the critical field is $<F_{crit}>$=14±4 kV/cm. As discussed in section 2.2, $F_{crit}$ is most sensitive to the difference of the electronic structure of the upper and lower dots. Indeed, for the model QD-molecule parameters used to fit the data of Figure 2.5 we assumed slightly detuned QDs. Comparing the distribution of $F_{crit}$ with the associated spectral detuning $\Delta E$ of the QDs (arrows in Figure 2.6c) we find an $<\Delta E>=0\pm8$meV emphasising the good spectral overlap achieved between the excitonic states in two QD layers.

## 3.0 Outlook

We have summarised recent advances towards the development of coherent optoelectronic devices based on interband optical excitations in semiconductor quantum dot nanostructures. One of the underlying themes throughout the present paper is the need for selectivity and tunability in nanoscale systems – provided both by the use of selective *optical* manipulation of individual quantum states and electrically *tunable* quantum systems. It may be that other excitations in QDs, such as the spin of isolated charge carriers, eventually emerge to have the most favourable properties for quantum information processing with solid-state hardware. Indeed, it has been recently demonstrated that the electron spin degree of freedom is a particularly stable variable in QDs [62], which may even have coherence times approaching the microsecond range or even longer.[63] Furthermore, single spins can already be selectively generated [62] and detected [64] using optical techniques and, it may be possible to manipulate them over ultrafast timescales using pulsed laser sources.[65] Regardless of which qubit basis eventually emerges to have the most favourable properties, the ability to selectively initialize, control and readout the quantum state using optical means is likely to remain of paramount importance due to the unique selectivity that it provides. The rapid progress in the field over the past five years is a firm indication that viable systems for quantum information processing will emerge in the near future.

*Acknowledgements*

This work was financed by the Deutsche Forschungsgemeinschaft via **SFB631** (*Festkörperbasierte Quanteninformationsverarbeitung: Physikalische Konzepte und Materialaspekte*) and BMBF via *Förderschwerpunkt: Elektronenkorrelation und Dissipationsprozesse in Halbleiterquantenstrukturen* und *Förderschwerpunkt: nanoQUIT*

## **References**

1 D. Bouwmeester, A. Ekert, A. Zeilinger: The Physics of Quantum Information, (Springer, Berlin, 2000).
2 DI. I. Rabi: Space Quantization in a Gyrating Magnetic Field, Phys. Rev. **51**, 652 (1937).
3 T. H. Stievater, Xiaoqin Li, D. G. Steel, D. Gammon, D. S. Katzer, D. Park, C. Piermarocchi, L. J. Sham, Phys. Rev. Lett. **87**, 133603 (2001).
4 H. Kamada, H. Gotoh, J. Temmyo, T. Takagahara, H. Ando, Phys. Rev. Lett. **87**, 246401 (2001).
5 A. Zrenner, E. Beham, S. Stufler, F. Findeis, M. Bichler, G. Abstreiter, Nature **418**, 612 (2002).
6 H. Htoon, T. Takagahara, D. Kulik, O. Baklenov, A. L. Holmes, Jr., C. K. Shih, Phys. Rev. Lett. **88**, 087401 (2002).
7 P. Borri, W. Langbein, S. Schneider, U. Woggon, R. L. Sellin, D. Ouyang, D. Bimberg, Phys. Rev. B **66**, 081306 (2002).
8 L. Besombes, J. J. Baumberg, J. Motohisa, Phys. Rev. Lett. **90**, 257402 (2003).
9 P. Borri, W. Langbein, S. Schneider, U. Woggon, R. L. Sellin, D. Ouyang, D. Bimberg, Phys. Rev. Lett. **87**, 157401 (2001).
10 M. Bayer, A. Forchel, Phys. Rev. B **65**, 041308 (2002).
11 F. Findeis, M. Baier, E. Beham, A. Zrenner, G. Abstreiter, Appl. Phys. Lett. **78**, 2958 (2001).
12 W.-H. Chang et al., Phys. Rev. B **62**, 6959 (2000).
13 A. Patanè et al., Phys. Rev. B **62**, 11084 (2000).
14 P.W. Fry et al., Phys. Rev. Lett. **84**, 733 (2000).
15 E. Beham, A. Zrenner, and G. Böhm, Physica E **7**, 359 (2000).
16 D. Gammon, E. S. Snow, B. V. Shanabrook, D. S. Katzer, D. Park, Phys. Rev. Lett. **76**, 3005-3008 (1996).
17 M. Bayer, A. Kuther, A. Forchel, A. Gorbunov, V. B. Timofeev, F. Schafer, J. P. Reithmaier, T. L. Reinecke, S. N. Walck, Phys. Rev. Lett. **82**, 1748-1751 (1999).
18 Alexander Högele, Stefan Seidl, Martin Kroner, Khaled Karrai, Richard J. Warburton, Brian D. Gerardot, Pierre M. Petroff, Phys. Rev. Lett. **93**, 217401 (2004).
19 E. Beham, A. Zrenner, F. Findeis, M. Bichler, and G. Abstreiter, Appl. Phys. Lett. **79**, 2808 (2001).
20 S. Stufler, P. Ester, A. Zrenner, M. Bichler, Appl. Phys. Lett. **85**, 4202-4204 (2004).
21 E. Dekel, D. Gershoni, E. Ehrenfreund, D. Spektor, J. M. Garcia, and P. M. Petroff, Phys. Rev. Lett. **80**, 4991 (1998).
22 L. Landin, M. S. Miller, M.-E. Pistol, C. E. Pryor, and L. Samuelson, Science **280**, 262 (1998).
23 L. Allen, J. H. Eberly, *Optical resonance and two-level atoms* (Wiley, New York, 1975).
24 R. Oulton, A. I. Tartakovskii, A. Ebbens, J. Cahill, J. J. Finley, D. J. Mowbray, M. S. Skolnick, and M. Hopkinson, Phys. Rev. B **69**, 155323 (2004)
25 K. Brunner, G. Abstreiter, G. Böhm, G. Trankle, G. Weimann, Phys. Rev. Lett. **73**, 1138 (1994).
26 N. H. Bonadeo, G. Chen, D. Gammon, D. S. Katzer, D. Park, D. G. Steel, Phys. Rev. Lett. **81**, 2759 (1998).
27 P. Machnikowski, L. Jacak, Phys. Rev. B **69**, 193302 (2004).

28 P. Bianucci, A. Muller, C. K. Shih, Q. Q. Wang, Q. K. Xue, C. Piermarocchi, Phys. Rev. B **69**, 161303 (2004).
29 C. Piermarocchi, P. Chen, Y. S. Dale, and L. J. Sham, Phys. Rev. B **65**, 075307 (2002)
30 D. D. Di Vincenzo, Phys. Rev. A **51**, 1015 (1995), A. Barenco *et al., ibid*. **52**, 3457 (1995).
31 G. Chen *et al.*, Science **289**, 1906 (2000), X. Li *et al. ibid.* **301**, 809 (2003).
32 M. Bayer *et al*, Nature **405**, 923 (2000).
33 J. J. Finley *et al.,* Phys. Rev. B**63**, 073307 (2001).
34 F. Findeis *et al.*, Solid State Communications **114**, 227 (2000)
35 W. Wegscheider, G. Schedelbeck, G. Abstreiter, M. Rother, and M. Bichler, Phys. Rev. Lett. **79**, 1917 (1997), G. Schedelbeck, W. Wegscheider, M. Bichler and G. Abstreiter. Science, **278**, 1792, (1997)
36 G. S. Solomon, J. A. Trezza, A. F. Marshall, and J. S. Harris, Jr., Phys. Rev. Lett. **76**, 952, (1996)
37 G. Burkard, D. Loss, and D. P. DiVincenzo, Phys. Rev. B**59**, 2070 (1999).
38 B. Grandidier, Y. M. Niquet, B. Legrand, J. P. Nys, C. Priester, D. Stiévenard, J. M. Gérard, and V. Thierry-Mieg, Phys. Rev. Lett. **85**, 1068 (2000).
39 B. Legrand, B. Grandidier, J. P. Nys, D. Stiévenard, J. M. Gérard, and V. Thierry-Mieg, Appl. Phys. Lett. **73**, 96 (1998).
40 B. Legrand, J. P. Nys, B. Grandidier, J. P. Nys, D. Stiévenard, A. Lemaître, J. M. Gérard, and V. Thierry-Mieg, Appl. Phys. Lett. **74**, 2608, (1999).
41 Q. Xie, A. Madhukar, P. Chen, and N. P. Kobayashi, Phys. Rev. Lett. **75**, 2542, (1995).
42 M. Bayer *et al.*, Science **291**, 451 (2001)
43 K. Hinzer, M. Bayer, J. P. McCaffrey, P. Hawrylak, M. Korkunski, O. Stern, Z. R. Wasilewski, S. Fafard and A. Forchel. Phys. Stat. Sol. (b) **224**, 385, (2001)
44 See e.g. D. Bruls, P. M. Koenraad, H. W. M. Salemink, J. H. Wolter, M. Hopkinson and M. S. Skolnick. Appl. Phys. Lett. **82**, 3758, (2003)
45 S. Fafard, M. Spanner, J.P. McCaffrey, and Z.R. Wasilewski, Appl. Phys. Lett. **76**, 2268 (2000).
46 Z. R. Wasilewski, S. Fafard, and J.P. McCaffrey, J. Cryst. Growth. **201**, 1131 (1999).
47 G. Bester, J. Shumway and A. Zunger, Phys. Rev. Lett **93**, 047401, (2004)
48 I. Shtrichman, C. Metzner, B. D. Geradot, W. Y. Schoenfeld and P. M. Petroff, Phys. Rev. **B65**, 081303(R) (2002)
49 M. Sugisaki, H. W. Ren, S. V. Nair, K. Nishi and Y. Masumoto, Phys Rev **B66**, 235309 (2002)
50 H. J. Krenner, M. Sabathil, E. C. Clark, A. Kress, D. Schuh, M. Bichler, G. Abstreiter and J. J. Finley. Phys. Rev. Lett. **94**, 057402, (2005)
51 G. Ortner, M. Bayer, Y. Lyander-Geller, T. L. Reinecke, A. Kress, J. P. Reithmaier and A. Forchel. Phys. Rev. Lett. **94**, 157401, (2005)
52 E. Biolatti, R. C. Iotti, P. Zanardi and F. Rossi. Phys. Rev. Lett. **85**, 5647, (2000)
53 G. Bester and A. Zunger. cond-mat/0502184, Feb (2005)
54 J. M. Villas-Bôas, A. O. Govorov and S. E. Ulloa. Phys. Rev. B**69**, 125342 (2004).
55 J. Stangl, V. Holý, and G. Bauer, Rev. Mod. Phys. **76**, 725 (2004)
56 J. Tersoff *et al.*, Phys Rev. Lett. **76**, 1675 (1996), H. Heidemeier *et al., ibid.* **91**, 196103 (2003)

57 S. Fafard *et al.*, Phys. Rev. **B59**, 15368 (1999)
58 Q. Xie *et al.*Appl. Phys. Lett. **76**, 3082 (2000) H. Heidemeier *ibid*. **80**, 1544 (2002)
59 J. J. Finley *et al*. Phys. Rev. **B70**, 201308(R), (2004)
60 J. P. Perdew and A. Zunger. Phys. Rev. **B23**, 5048, (1981)
61 N. Liu, J. Tersoff, O. Baklenov, A. L. Holmes and C. K. Shih. Phys. Rev. Lett. **84**, 334, (2000)
62 M. Kroutvar, Y. Ducommun, D. Heiss, M. Bichler, D. Schuh, G. Abstreiter and J. J. Finley. Nature **432**, 81 (2004)
63 V. Cerletti, W A Coish, O. Gywat and D. Loss, Nanotechnology 16 , R27–R49, (2005)
64 F. Jelezko and J. Wrachtrup, J. Phys.: Condens. Matter **16**, R1089–R1104, (2004)
65 T. Calarco, A. Datta, P. Fedichev, E. Pazy and P. Zoller. Phys. Rev. A **68**, 012310 (2003)